\documentclass[aip,prb,amsmath,amssymb,preprint]{revtex4-2}

\usepackage{graphicx} % Needed for figures (if any)
\usepackage{dcolumn}% Align table columns on decimal point
\usepackage{bm}% bold math
\usepackage[T1]{fontenc}
\usepackage{mathptmx}
\usepackage{etoolbox}

\makeatletter
\def\@email#1#2{%
 \endgroup
 \patchcmd{\titleblock@produce}
  {\frontmatter@RRAPformat}
  {\frontmatter@RRAPformat{\produce@RRAP{*#1\href{mailto:#2}{#2}}}\frontmatter@RRAPformat}
  {}{}
}%
\makeatother

\begin{document}

%\preprint{AJP25-NT-00615}
% --- TITLE PAGE INFORMATION ---
% For double-anonymous review, author and affiliation information should be commented out.
\title[Is the Lorenz Gauge a Choice?]{Is the Lorenz Gauge a Choice?\\Gauge Freedom and the Structure of Electrodynamics}

% % % ===================================================================
%  SEÇÃO DE AUTORIA - AUTOR ÚNICO - VERSÃO FINAL OTIMIZADA
% ===================================================================

\author{Alexsandro Lucena Mota}
    
    % O comando \thanks{} cria uma nota de rodapé vinculada ao nome do autor.
    % É a maneira padrão e mais limpa de fornecer múltiplos pontos de contato
    % e designar a correspondência.
    \thanks{Corresponding author:\\[-1em] 
    \hspace{5em} Personal e-mail: \href{mailto:lucenalexster@gmail.com}{lucenalexster@gmail.com}. \\[-1em]
    \hspace{5em} Institutional e-mails: \href{mailto:lucena.alexsandro@discente.ufma.br}{lucena.alexsandro@discente.ufma.br}, 
    \href{mailto:alexsandro.mota@prof.edu.ma.gov.br}{alexsandro.mota@prof.edu.ma.gov.br}.}

% Afiliação primária, onde o trabalho foi desenvolvido.
\affiliation{Curso de Física Licenciatura, Departamento de Física, Universidade Federal do Maranhão (UFMA), São Luís, MA, Brazil}

% Afiliação secundária, que é pedagogicamente relevante.
% Opcional, mas recomendado para um artigo de ensino.
\altaffiliation[Also at: ]{Centro de Ensino Médio Y Bacanga, Secretaria de Estado da Educação (SEDUC), São Luís, MA, Brazil.}

\date{\today}

% --- ABSTRACT ---
\begin{abstract}
In undergraduate electromagnetism courses, the Lorenz gauge condition is often presented as a convenient mathematical choice that decouples the wave equations for the scalar and vector potentials. While true, this presentation may leave students with the impression that the condition is entirely arbitrary. This Note explores the fundamental structure of gauge invariance, demonstrating that the Lorenz condition is not an ad-hoc imposition but rather the most elegant and natural simplification afforded by the theory's inherent gauge freedom. We explicitly show how the gauge function itself transforms, proving that one can always choose a gauge in which the Lorenz condition holds. This approach aims to transform the topic from a formal trick into an instructive example of the structure of gauge theories.
\end{abstract}

\maketitle

% --- ARTICLE BODY ---

\section{\label{sec:level1}Introduction}

Introducing the scalar potential $\phi$ and the vector potential $\boldsymbol{A}$ is a cornerstone of classical electrodynamics. They simplify Maxwell's equations by satisfying, by construction, the two homogeneous equations: Gauss's law for magnetism ($\nabla \cdot \boldsymbol{B} = 0$) and Faraday's law ($\nabla \times \boldsymbol{E} = -\partial\boldsymbol{B}/\partial t$).\cite{Jackson_Book} This formalism, however, leads to a system of coupled, second-order differential equations for the potentials. The canonical textbook approach, prominently exemplified in Chap.~10 of Griffiths, is to invoke the invariance of the fields under gauge transformations to impose the Lorenz gauge condition,
\begin{equation}
    \nabla \cdot \boldsymbol{A} + \frac{1}{c^2} \frac{\partial \phi}{\partial t} = 0.
    \label{eq:lorenz_condition}
\end{equation}
As Griffiths notes, this choice is specifically ``designed to eliminate the middle term'' in the original coupled wave equation for $\boldsymbol{A}$. This imposition indeed decouples the equations, leading to the elegant result where both potentials satisfy the same differential operator. In Griffiths' words, ``The virtue of the Lorenz gauge is that it treats $\phi$ and $\boldsymbol{A}$ on an equal footing.'' (Ref.~\onlinecite{Griffiths_Book}, p.~441)

While this mathematical elegance is powerful, the presentation can be pedagogically unsatisfying. It frames a condition of profound physical importance as a mere matter of mathematical convenience.  This type of pedagogical simplification has long been a subject of critical discussion in the physics education literature, for instance, in the context of displacement currents.\cite{French_Tessman} This leaves a persistent question: Is the Lorenz\cite{Jackson_Okun_RMP} condition \textit{just} a convenient choice, or is there a deeper physical reason for its specific form? Discussions on the subtleties of gauge choice, while central to the theory, often remain secondary in undergraduate curricula. Indeed, as Jackson noted in this Journal, ``textbooks rarely show explicitly the gauge function $\chi$ that transforms one gauge into another,'' leaving students without the tools to explore the structure of gauge freedom for themselves.\cite{Jackson_Gauge_AJP} Other authors have also highlighted how dynamical and consistency aspects of the potentials in various gauges are often omitted from the standard treatment.\cite{Yang_Gauge, Heras_Gauge}

The purpose of this Note is to investigate precisely this question by focusing on the structure of gauge freedom itself. We will demonstrate that the Lorenz condition emerges as the most natural and convenient choice allowed by the theory's structure, without needing to be constrained by other physical principles like charge conservation, which must be satisfied regardless. This approach aims to transform the topic of gauge freedom from a formal trick into an instructive example of the mathematical structure of our physical theories.

\section{The Dynamics of the Gauge Condition}

The dynamics of the electromagnetic potentials, prior to the imposition of a gauge condition, are described by the coupled equations derived from the inhomogeneous Maxwell's equations.\cite{Jackson_Book} To explore the scope of gauge freedom explicitly, we posit a generalized gauge condition by defining an, in principle, arbitrary function $f(\boldsymbol{r}, t)$:
\begin{equation}
    \nabla \cdot \boldsymbol{A} + \frac{1}{c^2} \frac{\partial \phi}{\partial t} = f(\boldsymbol{r}, t).
    \label{eq:general_f}
\end{equation}
The standard Lorenz gauge corresponds to the choice $f=0$. Substituting this generalized condition into the coupled equations for the potentials yields a decoupled system of inhomogeneous wave equations:
\begin{align}
    \Box \phi &= \frac{\rho}{\epsilon_0} + \frac{\partial f}{\partial t}, \label{eq:wave_phi_f} \\
    \Box \boldsymbol{A} &= \mu_0 \boldsymbol{J} - \nabla f, \label{eq:wave_A_f}
\end{align}
where we have adopted the d'Alembertian operator definition $$\Box \equiv \frac{1}{c^2}\frac{\partial^2}{\partial t^2} - \nabla^2.$$

To understand the nature of the function $f$, we must first recall that the physical fields $\boldsymbol{E}$ and $\boldsymbol{B}$ must remain invariant under any choice of gauge. We know that the potentials are not unique; a gauge transformation given by
\begin{align}
    \boldsymbol{A}' &= \boldsymbol{A} + \nabla\chi, \\
    \phi' &= \phi - \frac{\partial\chi}{\partial t},
\end{align}
for any arbitrary scalar function $\chi(\boldsymbol{r}, t)$, leaves the physical fields unchanged. Let us now apply this transformation to our generalized gauge condition, Eq.~\eqref{eq:general_f}. The new gauge function, $f'$, is defined by
\begin{equation}
    f' = \nabla \cdot \boldsymbol{A}' + \frac{1}{c^2} \frac{\partial \phi'}{\partial t}.
\end{equation}
Substituting the transformed potentials, we find how the new gauge function $f'$ relates to the original function $f$:
\begin{align}
    f' &= \nabla \cdot (\boldsymbol{A} + \nabla\chi) + \frac{1}{c^2} \frac{\partial}{\partial t} \left(\phi - \frac{\partial\chi}{\partial t}\right) \nonumber \\
       &= \left(\nabla \cdot \boldsymbol{A} + \frac{1}{c^2}\frac{\partial\phi}{\partial t}\right) - \left(\frac{1}{c^2}\frac{\partial^2\chi}{\partial t^2} - \nabla^2\chi\right) \nonumber \\
       &= f - \Box\chi.
       \label{eq:transform_f}
\end{align}
This equation is the central result of our analysis. It proves that the function $f$ is not a gauge-invariant physical quantity. Its form depends entirely on the gauge choice we make, which is parameterized by the function $\chi$. This transformation property, while fundamental, is typically explored in more advanced treatments of the subject.\cite{Jackson_Book}

The second, and most crucial, point is that we can use this very gauge freedom to simplify our theory without any loss of physical content. Eq.~\eqref{eq:transform_f} shows that, given any non-zero function $f$ resulting from an initial choice of potentials, we can always seek a gauge transformation function $\chi$ that satisfies the inhomogeneous wave equation:
\begin{equation}
    \Box\chi = f.
    \label{eq:wave_chi}
\end{equation}
The existence of solutions to the inhomogeneous wave equation is a standard result in mathematical physics, typically found by employing Green's functions.\cite{Arfken_Book} Its specific application to electrodynamics is a cornerstone of advanced treatments of the subject.\cite{Jackson_Book} This guarantees that we can always perform a gauge transformation to a new gauge where the new function $f'$ becomes zero.

This approach demonstrates that the Lorenz condition is not an arbitrary ``imposition'' on the theory, but rather the most elegant and convenient simplification that emerges naturally from the inherent gauge freedom of electromagnetism. It allows the potentials to be described by the simplest possible decoupled wave equations, making calculations far more direct, without losing any physical information.

\section{Conclusion}

The argument presented clarifies the nature of gauge freedom: it is the liberty to redefine the potentials in a way that simplifies the mathematical description of a system. We have shown that for any set of potentials, a gauge transformation can always be found that enforces the Lorenz condition. The condition $f=0$ is not a law of nature, but a convention --- a particularly wise one that simplifies the equations of motion to their minimal, decoupled form. As Jackson summarizes, the Lorenz gauge is commonly used because it ``treat[s] $\Phi$ and $\boldsymbol{A}$ on equivalent footings'' and ``fits naturally into the considerations of special relativity'' (Ref.~\onlinecite{Jackson_Book}, p.~241).

Furthermore, the fact that $f$ can be transformed away confirms that it does not correspond to a measurable quantity. The Lorenz condition is thus the choice that eliminates this unphysical redundancy in the simplest, most elegant, and relativistically covariant manner. Presenting this argument in the classroom provides valuable insight into the structure of gauge theories, showing how conservation laws and symmetries shape the mathematical language of physics.

% --- BIBLIOGRAPHY (Manual entries for RevTeX) ---
% The numeral (99) nominally sets the width of the labels.

\end{document}